\documentclass[aps,onecolumn,amsmath,amssymb,superscriptaddress,floatfix,nofootinbib]{revtex4}
\usepackage{graphicx}
\usepackage{epsfig}
\usepackage{epstopdf}
\usepackage{multirow}
\usepackage{color}
\usepackage[colorlinks=true,urlcolor=blue,linkcolor=blue,citecolor=blue]{hyperref}
\usepackage{hyperref}
\usepackage{amsmath}
\usepackage{amsfonts}
\usepackage{amssymb}
\usepackage{bbold}
\usepackage{braket}
\usepackage{xparse}
\usepackage{makecell}

\def \) {\right)}
\newcommand{\abs}[1]{\left\lvert#1\right\rvert}

\begin{document}

\title{Lattice Improvement in Lattice Effective Field Theory}

\author{Nico~Klein} 
\affiliation{Helmholtz-Institut f\"ur Strahlen- und
             Kernphysik and Bethe Center for Theoretical Physics,
             Universit\"at Bonn,  D-53115 Bonn, Germany}  
             
\author{Dean Lee}                 
\affiliation{National Superconducting Cyclotron Laboratory, Michigan State University, East Lansing, MI 48824, USA}
\affiliation{Department of Physics, North Carolina State University, Raleigh, NC 27695, USA}

\author{Ulf-G.~Mei{\ss}ner}  
\affiliation{Helmholtz-Institut f\"ur Strahlen- und
             Kernphysik and Bethe Center for Theoretical Physics,
             Universit\"at Bonn,  D-53115 Bonn, Germany}  
\affiliation{Institute~for~Advanced~Simulation, Institut~f\"{u}r~Kernphysik,
and J\"{u}lich~Center~for~Hadron~Physics,~Forschungszentrum~J\"{u}lich,
D-52425~J\"{u}lich, Germany}     
\affiliation{JARA~-~High~Performance~Computing, Forschungszentrum~J\"{u}lich,
D-52425 J\"{u}lich,~Germany} 
\date{\today}
\begin{abstract}
Lattice calculations using the framework of effective field theory have been applied to a wide range few-body and many-body systems.  One of the challenges of these calculations is to remove systematic errors arising from  the nonzero lattice spacing. Fortunately, the lattice improvement program pioneered by Symanzik provides a formalism for doing this. While
lattice improvement has already been utilized in lattice effective
field theory calculations, the effectiveness of the improvement program has not been systematically benchmarked. In
this work we use lattice improvement to remove lattice errors for a one-dimensional
system of bosons with zero-range interactions.  We construct the improved lattice action up to next-to-next-to-leading order and verify that the remaining errors scale as the fourth power of the lattice spacing for observables involving as many as five particles.
Our results provide a guide for increasing the accuracy of future calculations in lattice effective field theory with improved lattice actions.  
\end{abstract}

\maketitle
\section{Introduction}
Lattice
simulations based on the framework of effective field theory (EFT) have been used in the study of nuclear forces \cite{Alarcon:2017zcv,Li:2018ymw}, nuclear structure \cite{Epelbaum:2011md,Epelbaum:2012qn,Lahde:2013uqa,
Epelbaum:2013paa,Elhatisari:2016owd,Elhatisari:2017eno}, as well as scattering and 
reactions \cite{Rupak:2013aue,Elhatisari:2015iga,Elhatisari:2016hby}.  It has also been applied to  few-body \cite{Bour:2012hn,Elhatisari:2016hui} 
and many-body \cite{Bulgac:2005pj,Lee:2008xsa,Carlson:2011kv,Endres:2012cw,Bour:2014bxa,
Braun:2014pka,Anderson:2015uqa} problems in ultracold atomic systems.  One of the challenges common to all of these lattice calculations is the need to eliminate errors caused by the non-vanishing lattice spacing $a$.
Lattice spacing errors or artifacts can grow in importance as the number of particles increases. This is especially true in cases where the particles form compact bound states. The problem of lattice errors in computing compact bound states has been studied in two-dimensional droplets of bosons with attractive zero range interactions \cite{Lee:2005nm}.  In Ref.~\cite{Klein:2018lqz} lattice spacing errors were studied in light nuclei to understand deviations from a universal relation  between the triton and alpha-particle binding energies called the Tjon line \cite{Tjon:1975sme,Platter:2004zs}.

One practical approach to reducing lattice artifacts is to introduce a continuous-space regulator that renders the particle interactions ultraviolet finite.  
In that case one can take the lattice spacing to zero smoothly 
without any need to renormalize operator coefficients \cite{Montvay:2012zz,Klein:2015vna}.  This process does not remove errors produced by the 
continuous-space regulator, but it has the advantage of eliminating unphysical lattice artifacts such as rotational symmetry breaking. 
Another approach to removing lattice artifacts is accelerating the convergence to the continuum limit by including irrelevant higher-dimensional 
operators into the lattice action. This general technique is called lattice improvement and was first introduced by Symanzik \cite{Symanzik:1983dc,Symanzik:1983gh}.
The key idea is to write the lattice action as    
\begin{equation}
S(a) = S_0 + f_1(a)\cdot S_1 + f_2(a)\cdot S_2 + \cdots 
\end{equation}   
where the coefficient functions $f_i(a)$ vanish in the continuum limit $a\rightarrow \infty$   and are tuned to cancel the dependence of 
the lattice Green's functions on the lattice spacing, $a$, at momentum scales well below the lattice cutoff momentum, $\pi/a$. 
The coefficient functions are ordered so that $f_i(a)$ vanishes more rapidly in the limit $a\rightarrow 0$ with increasing  $i$.  

Symanzik's improvement program has been widely used in lattice quantum chromodynamics \cite{Luscher:1984xn,Sheikholeslami:1985ij,Luscher:1996sc,Luscher:1996ug}.  
The improvement program has also been implemented to varying degrees in lattice effective field theory calculations.  However, a systematic study of the 
size of lattice errors arising in systems with increasing numbers of particles has not yet been performed. In this work we address this problem and 
benchmark the effectiveness of the lattice improvement program in removing lattice errors for a one-dimensional
system of bosons with zero-range interactions \cite{McGuire:1964zt}. By building the improved lattice action step by step, we show that the 
lattice errors go from $\mathcal{O}(a^1)$ at leading order (LO), to $\mathcal{O}(a^3)$ at next-to-leading order (NLO), and then to $\mathcal{O}(a^4)$ 
at next-to-next-to-leading order (N2LO). Our discussion is organized as follows. In Sec.~\ref{sec:theory} we introduce the one-dimensional 
boson system in the continuum, and in Sec.~\ref{sec:lattice} we present the lattice implementation.  We show and discuss our results in 
Sec.~\ref{sec:results} and then summarize and give an outlook for future applications in Sec.~\ref{sec:summary}.

\section{Bosons in one dimension}\label{sec:theory}

We consider a one-dimensional system of bosons with attractive zero-range interactions \cite{McGuire:1964zt}. The Hamiltonian is given by
\begin{equation}
H = -\frac{1}{2m}\sum_{i=1}^N \frac{d^2}{dx_i^2}+C_0\sum_{i>j=1}^N\delta(x_i-x_j),
\label{eq:Lcontinuum}
\end{equation}
where $m$ is the boson mass, $N$ is the number of particles and $C_0<0$ is the two-boson contact interaction coefficient. From either the 
Bethe ansatz \cite{Bethe:1931hc} or by direct inspection, the ground state of this Hamiltonian is a bound state  with energy 
\begin{equation}
E_{B}^{N\mathrm{b}} = -\frac{m}{24}C_0^2 N(N^2-1),
\end{equation}
and wave function
\begin{equation}
\Psi_N\left(x_1,x_2,\ldots,x_N\right)  = \mathcal{N} \exp \left[\frac{mC_0}{2}\left(\abs{x_1-x_2}+\abs{x_1-x_3}+\ldots\right)\right],\label{eq:Psiexact}
\end{equation}
with normalization constant $\mathcal{N}$ \cite{McGuire:1964zt}.  We note that no regularization or renormalization is needed, and all quantities are finite. 

In this work we consider several different bound state and scattering observables.  We will consider the binding energies of the $N$-boson bound 
state for up to five bosons and the root-mean-square radii of their corresponding wave functions. The root-mean-square radii for all the particle 
positions will be the same due to Bose symmetry and can be  calculated as
\begin{equation}
r_{\rm rms}^2=\int dx_1 \cdots dx_N \abs{x_1-\frac{x_1+\cdots+x_N}{N}}^2\left|\Psi_N\left(x_1,x_2,\ldots,x_N\right)\right|^2 .
\end{equation}
For the scattering observables, we compute coefficients of the effective range expansion for the boson-boson and dimer-boson even-parity scattering 
phase shifts, which we denote as $\delta_2$ and $\delta_3$, respectively.  We use the conventions 
\begin{align}
p\tan\delta_2 &=\frac{1}{a_2}+\frac{r_2}{2}p^2+\ldots \, , \label{r2}\\ 
p\tan\delta_3 &=\frac{1}{a_3}+\frac{r_3}{2}p^2+\ldots, \,  
\end{align}
where $a_{i}$ is the scattering length, $r_{i}$ is the effective range, and $p$ is the relative momentum. From the Bethe ansatz, these effective range functions are known exactly \cite{Mehta:2005nq},
\begin{align}
p \tan\delta_2 &=\sqrt{-m E_{B}^{2\mathrm{b}}} \, ,\\
p \tan\delta_3 &=-4\frac{a_2p^2}{\frac{3}{2}a_2^2p^2-2}\, ,\label{eq:ptan3}
\end{align}
where $E_{B}^{2\mathrm{b}}$ is the two-boson bound state energy and  $a_2=-2/(mC_0)$ is the two-boson scattering length.

When implemented on a spatial lattice, this theory will be modified by the lattice momentum cutoff scale $\Lambda=\pi/a$. In order to study the 
size of the lattice errors, we must know the dependence of the Green's functions upon the cutoff momentum.  For this analysis we use perturbation 
theory and determine the cutoff dependence of individual Feynman diagrams.  In some cases there can be non-perturbative effects that arise from 
correlations that modify the ultraviolet properties of the Green's functions. However, perturbation theory is still a good starting point, and 
the cutoff error estimates  can then be verified by numerical calculations.  For our one-dimensional system of bosons, all diagrams are 
ultraviolet finite and so  all of the cutoff effects will be of residual dependences that vanish in the continuum limit.

\begin{figure}[!t]
\centering
\includegraphics[width=0.25\textwidth]{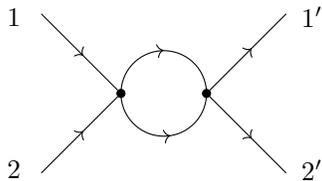}
\caption{One-loop diagram $\mathcal{I}_2$ involving two bosons in the initial and the final state.}
\label{I2}
\centering
\end{figure}

The Feynman diagram with the highest degree of divergence (albeit negative) is the one-loop diagram $\mathcal{I}_2$ involving two bosons as shown in Fig.~\ref{I2}.
Let the energies and momenta of the incoming particles be
$(E_1,p_1)$ and $(E_2,p_2)$, and the energies and momenta of the outgoing particles be
$(E'_1,p'_1)$ and $(E'_2,p'_2)$.
We write $\Delta I_2$ for the difference between the continuum-limit amplitudes and the lattice amplitudes at lattice spacing $a$.   
We expand in powers of momenta and energies and get  
\begin{equation}
\Delta I_2 =  A_2(a) + (p_1+p_2)^2B_2(a) + m(E_1+E_2)C_2(a) + \mathcal{O}(p^4), \label{eq:I2diag}
\end{equation} 
where $\mathcal{O}(p^4)$ is shorthand for terms with more powers of momentum or energy.  From dimensional analysis we find that $A_2(a) \sim \mathcal{O}(a^1)$, 
while $B_2(a) \sim \mathcal{O}(a^3)$ and $C_2(a) \sim \mathcal{O}(a^3)$.  Terms with higher powers of momentum or energy start at $\mathcal{O}(a^5)$. 
The dependence on the total momentum, $p_1+p_2$, is a result of the fact that the lattice regulator breaks Galilean invariance and so the residual 
amplitude $\Delta I_2$ can depend on the frame of reference.

\begin{figure}[!t]
\centering
\includegraphics[width=0.25\textwidth]{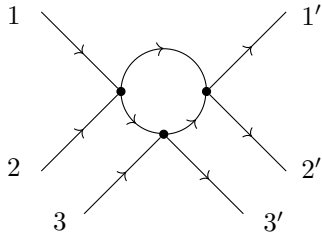}
\caption{One-loop diagram $\mathcal{I}_3$ involving three bosons in the initial and the final state.}
\label{I3}
\centering
\end{figure}

The diagram with the next highest degree of divergence
is the one-loop diagram $\mathcal{I}_3$ involving three bosons as shown in Fig.~\ref{I3}.
Let the energies and momenta of the incoming particles be
$(E_1,p_1)$, $(E_2,p_2)$, $(E_3,p_3)$ and the energies and momenta of the outgoing
particles be
$(E'_1,p'_1)$, $(E'_2,p'_2)$, $(E'_3,p'_3)$.
We let $\Delta I_3$ be the difference between the continuum-limit and lattice amplitudes.
Expanding in powers of momenta and energies gives 
\begin{equation}
  \Delta I_3 =  A_3(a) + \mathcal{O}(p^2).\label{eq:I3diag}
\end{equation}
From dimensional analysis we conclude that $A_3(a) \sim \mathcal{O}(a^3)$ and the terms with higher powers of momentum or energy start 
at $\mathcal{O}(a^5)$.  Our analysis thus far might give the false impression that all other errors start at $\mathcal{O}(a^5)$.  
However, we should note that there are two-loop diagrams involving three bosons that give a cutoff dependence that is $\mathcal{O}(a^4)$.  
Furthermore, there will be $\mathcal{O}(a^4)$ errors in the two-boson system when we iterate the improved energy-independent interactions that 
we use to cancel the $\mathcal{O}(a^1)$ and $\mathcal{O}(a^3)$ errors. We give further details for the expressions appearing in 
Eq.~(\ref{eq:I2diag}) and Eq.~(\ref{eq:I3diag}) in the Appendix~\ref{app:loops}.

\begin{figure}[!t]
\centering
\includegraphics[width=0.25\textwidth]{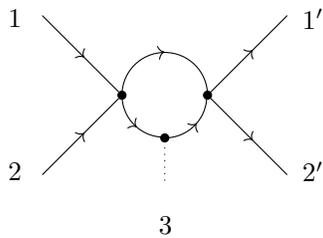}
\centering
\caption{One-loop form factor diagram $\mathcal{I}'_2$ involving two bosons in the initial and the final state and one external insertion. }
\label{Ip2}
\end{figure}

Since we are computing the root-mean-square radius, we also need to consider form factor diagrams associated with the boson density.  
The form factor diagram with the highest degree of divergence is the one-loop diagram $\mathcal{I}'_2$ shown in Fig.~\ref{Ip2}.  
Structurally this diagram is the same as $\mathcal{I}_3$, and so the residual amplitude will have a similar form,
\begin{equation}
  \Delta I'_2 =  A'_2(a) + \mathcal{O}(p^2).
\end{equation}
From dimensional analysis $A'_2(a)$ will be $\mathcal{O}(a^3)$.  In this case, however, the momentum-independent part of the form factor just counts 
the number of bosons. So $A'_2(a)$ will vanish if number conservation is properly implemented on the lattice, and we assume this is the case. 
The root-mean-square radius is proportional to the derivative of the form factor with respect to the squared momentum transfer, and so 
the correction to the root-mean-square radius from this diagram will be $\mathcal{O}(a^5)$.

\section{Lattice implementation}\label{sec:lattice}

We now implement our one-dimensional theory of bosons on the lattice. We use lattice units where all quantities are multiplied
by powers of the lattice spacing to form dimensionless combinations.  Since we are interested in methods that
can be readily applied to many-body calculations, we work with energy-independent interactions only.  We
use an $\mathcal{O}(a^4)$-improved free Hamiltonian lattice action,
\begin{equation}
H_{\rm free}=\frac{1}{2 m}\sum_{n}2 \omega_0 a^\dagger (n)a (n)  
+\frac{1}{2 m}\sum_{l=1}^3(-1)^l\omega_l\left[a^\dagger(n+l)a(n)+a^\dagger(n)a (n+l)\right],
\label{eq:Hfree}
\end{equation}
with coefficients
\begin{equation}
\omega_0= \frac{49}{36},\quad\omega_1 = \frac{3}{2},\quad \omega_2 = \frac{3}{20},\quad \omega_3 = \frac{1}{90}\, .
\end{equation}
The lattice artifacts due to the free Hamiltonian start at $\mathcal{O}(a^6)$.
At LO, we have a contact interaction in the two-boson sector,
\begin{equation}
V_{\rm LO}=\frac{C_0}{2} \sum_{n} \colon \rho^2(n) \colon, 
\label{eq:HLOcontact}
\end{equation}
where $\rho(n)=a^\dagger(n)a(n)$ and the $::$ symbols indicate normal ordering where the annihilation operators stand to the right  
and the creation operators stand to the left. The lattice contact interaction is tuned to the same value as the continuum-limit 
interaction, $C_0$.  At NLO we have a finite renormalization of the LO interaction,
\begin{equation}
V_{\rm NLO}=\frac{\Delta C_0}{2} \sum_{n} \colon \rho^2(n) \colon,\label{eq:HNLOcontact}
\end{equation}
which is sufficient to cancel the function $A_2(a)$ in Eq.~(\ref{eq:I2diag}).

At N2LO, two additional operators are required in the two-boson sector. The first is proportional to the square of the 
transferred momentum between particles,
\begin{equation}
V_{{\rm N2LO},q^2}^{\rm } = -\frac{C_{q^2}}{2} :\sum_{n}  \rho(n)
\left[ \rho(n+1) +\rho(n-1)-2\rho(n) \right] :,
\end{equation}and it removes the residual energy dependence $C_2(a)$ in Eq.~(\ref{eq:I2diag}) for on-shell two-boson scattering. 
The second N2LO operator is a Galilean-invariance-restoring (GIR) term that is proportional to the square of the total momentum,
\begin{align}
 V_{\rm N2LO,GIR}^{\rm } &= -\frac{C_{\rm GIR}}{8}\sum_{n}
 \left[a^\dagger(n+1)a^\dagger(n+1)a(n)a(n)+a^\dagger(n-1)a^\dagger(n-1)a(n)a(n)-2a^\dagger(n)a^\dagger(n)a(n)a(n) \right].\label{eq:gsb2N}
\end{align}
We use this operator to correct the two-boson dispersion relation on the lattice, thereby removing
the $B_2(a)$ term in Eq.~(\ref{eq:I2diag}).
One three-boson interaction is also needed at order N2LO,
\begin{equation}
V^{\rm }_{\rm N2LO,3b} = \frac{C_{\rm 3b}}{6}\sum_{n}\colon\rho^3
\left(n\right)\colon,\\
\end{equation}
which cancels the contribution the function $A_3(a)$ in Eq.~(\ref{eq:I3diag}). 
In all of our lattice calculations, we must also careful that finite volume errors are not being confused with lattice discretization errors.  
To this end, we always take the volume to be sufficiently large so that the finite volume errors are negligible.

\section{Results}\label{sec:results}

We now present lattice results at LO, NLO and N2LO and the discrepancies that remain when compared with the zero-range continuum limit. All continuum 
observables are calculated with $C_0=-0.1$ and $m=938.92$~MeV, which give binding energies roughly comparable to that of the deuteron, 
triton and helium-4. Of course, in the real world there is no five-body nucleonic state analogous to the five-boson bound state we 
consider here. Nevertheless, our analysis of residual lattice errors for this bosonic system will demonstrate the steps needed for 
systematic lattice improvement in nuclear lattice simulations.

We vary the lattice spacing from a maximum of $a=1.97$~fm to values as small as computationally possible, while keeping the physical
box length large enough so that finite volume effects are negligible.
The determination of the lattice parameters are as follows.  At LO, we consider the contact interaction with the continuum value $C_0=-0.1$, and 
no fitting is needed.   At NLO we fit $\Delta C_0$ to reproduce the two-boson continuum bound-state energy $E_{B}^{2\mathrm{b}}$.  At N2LO, 
we fit the coefficients $\Delta C_0$, $C_{q^2}$, $C_{\rm{GIR}}$ and $C_{\rm 3b}$ in order to simultaneously reproduce the continuum two-boson 
bound-state energy $E_{B}^{2\mathrm{b}}$, two-boson effective range $r_2$, two-boson effective mass $m_{\rm eff,2}$, and the three-boson bound-state 
energy $E_{B}^{3\mathrm{b}}$. We determine the effective range using Eq.~(\ref{r2}) and compute the elastic scattering phase
shifts at finite volume using the relation~\cite{Luscher:1986pf},
\begin{equation}
\exp\left[2 i \delta(p)\right] \exp\left(ipL\right) = 1~.
\end{equation}

\begin{table}[!t]
\begin{tabular}{|c|c |c| c| c |c|}
\hline\hline
row & observable & continuum result & LO & NLO & N2LO \\
\hline
1 & $E_{B}^{2\mathrm{b}}$     & $-2.347$ MeV &0.974(6) & --- & --- \\
\hline
2 &$r_{\rm rms}^{2\mathrm{b}}$ & 1.486 fm & 0.997(3)&3.05(2) &4.1(3) \\
\hline
3 &$E_{B}^{3\mathrm{b}}$ & $-9.389$ MeV & 0.97(2)& 2.97(2)& ---\\
\hline
4 &$r_{\rm rms}^{3\mathrm{b}}$ & 1.108 fm & 0.97(2)& 3.01(1)& 4.0(3)\\
\hline
5 &$E_{B}^{4\mathrm{b}}$ & $-23.473$ MeV & 0.967(5)& 3.05(1) & 3.9(1)\\
\hline
6 &$r_{\rm rms}^{4\mathrm{b}}$ & 0.867 fm & 0.97(1) & 3.09(5) & 4.1(2)\\
\hline
7 &$E_{B}^{5\mathrm{b}}$ & $-46.946$ MeV & 0.94(2) & 3.21(4) & 3.9(2)\\
\hline
8 &$r_{\rm rms}^{5\mathrm{b}}$ & 0.709 fm & 0.86(5) & 3.4(1) & 4.0(2)\\
\hline
9 &$m_{\rm eff}^{2\mathrm{b}}$ &1877.8 MeV & 2.91(9) & 2.96(5) & ---\\
\hline
10 &$1/a_{2}$ & 46.946 MeV & 0.978(6)& 2.99(6)& ---\\
\hline
11 &$1/a_{3}$ & 0 MeV & 2.80(3)& 2.93(6)& 4.08(3)\\
\hline\hline
\end{tabular}
\caption{Continuum observables and lattice error exponents $b_O$ as defined in Eq.~(\ref{eq:extrap}).}
\label{tab:obssum}
\end{table}

\begin{table}[!t]
\begin{tabular}{|l |c| c| c |c|c|}
\hline\hline
& NLO & \multicolumn{4}{c|}{N2LO}\\
\hline
$a$ [fm] & \makecell{$\Delta C_0$ \\$[\times 10^{-3}]$}& \makecell{$\Delta
C_0$ \\$[\times 10^{-3}]$}& \makecell{$C_{\rm 3b}$ \\$[\times 10^{-3}]$} & \makecell{$C_{q^2}$\\
$[\times 10^{-4}]$} & \makecell{$C_{\rm GIR}$ \\ $[\times 10^{-3}]$} \\
\hline
1.973  &$-8.79793$ & $-9.09089$ & 1.04056 & 7.16271 & 1.43521 \\
1.715  &$-7.55226$ & $-7.72824$ & 1.11585 & 4.55222 & 1.23930 \\
1.315  &$-5.67790$ & $-5.75996$ & 0.97610 & 2.41371 & 0.93898 \\
0.986  &$-4.19286$ & $-4.23347$ & 0.64741 & 1.41280 & 0.69463 \\
0.789  &$-3.32409$ & $-$3.34855 & 0.43540 & 0.98669 & 0.54967 \\
0.659  &$-2.75375$ & $-$2.77014 & 0.30594 & 0.75364 & 0.45367 \\
0.563  &$-2.35055$ & $-$2.36231 & 0.22429 & 0.60773 & 0.38527 \\
0.438  &$-1.81824$ & $-$1.82514 & 0.13387 & 0.43622 & 0.29366 \\
0.373  &$-1.53939$ & $-$1.54429 & 0.09575 & 0.35513 & 0.24475 \\
0.303  &$-1.25151$ & $-$1.25473 & 0.06429 & 0.27726 & 0.19161 \\
0.282  &$-1.16105$ & $-$1.16381 & 0.05604 & 0.25397 & 0.17423 \\
0.246  &$-1.01441$ & $-$1.01650 & 0.04444 & 0.21731 & 0.14466 \\
0.164  &$-0.67394$ & $-$0.67486 & 0.02664 & 0.13771 & 0.06049 \\
0.098  &$-0.40326$ & $-$0.40360 & 0.02427 & 0.08263 & 0.03280 \\
\hline\hline
\end{tabular}
\caption{Low-energy coupling constants for NLO and N2LO. All constants are
in lattice units.}
\label{tab:LECs}
\end{table}

\begin{figure}[!t]
\center
\includegraphics[width=1.0\textwidth]{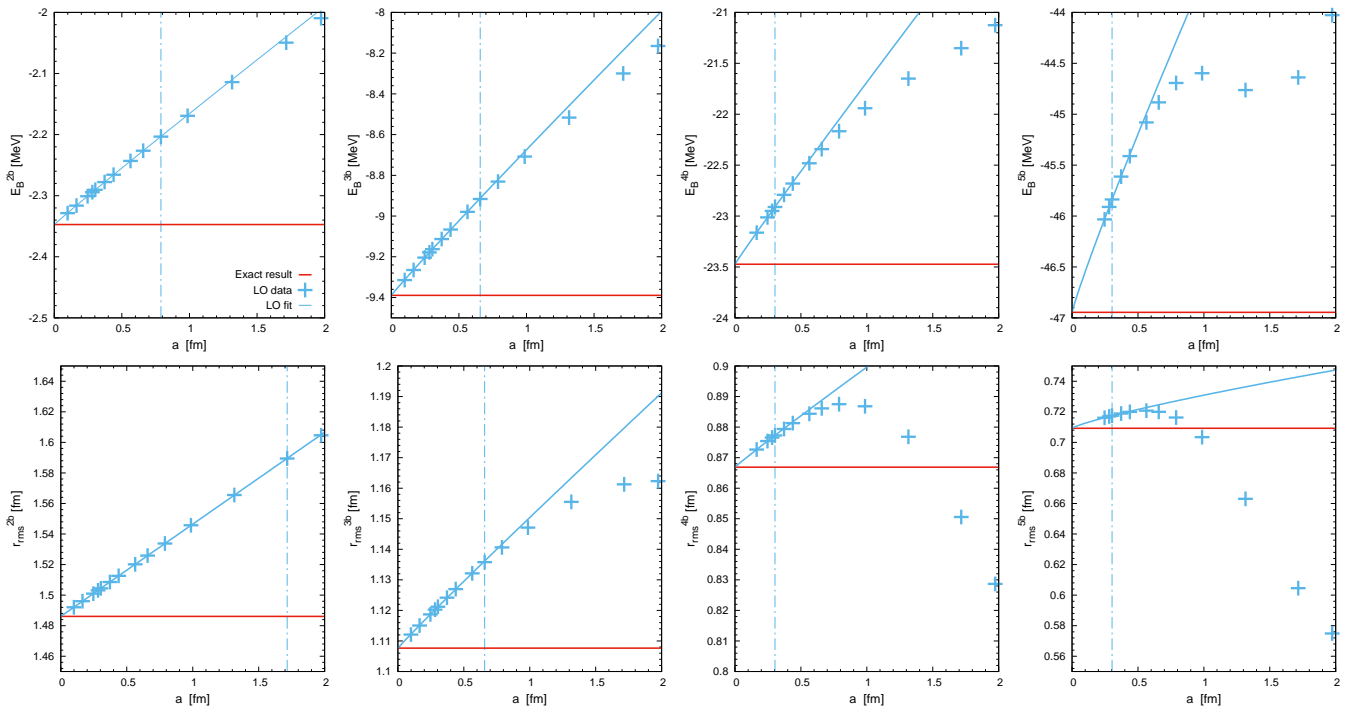}
\caption{Bound-state energies and root-mean-square radius versus lattice spacing for up to five bosons at LO. The vertical lines 
give the upper limits of the fit range.}
\label{fig:energyrmsLO}
\end{figure}

Each lattice observable is fitted using the asymptotic fit function, 
\begin{equation}
O(a)=O_c+A_O\cdot a^{b_O},
\label{eq:extrap}
\end{equation}
where $O_c$ is the zero-range continuum-limit value of the observable and $A_O$ and $b_O$ are fit parameters in the limit $a\to 0$. 
We show the LO bound state energies and root-mean-square radii for up to five bosons in Fig.~\ref{fig:energyrmsLO} as well as the best fits 
using the asymptotic form in Eq.~(\ref{eq:extrap}). The vertical lines give the
upper limits of the fit range.  As we see from rows one through eight of Table~\ref{tab:obssum}, 
the LO exponent $b_O$ is consistent with 1. This is the same as the $\mathcal{O}(a^1)$ estimate we derived from our perturbative analysis 
and the $A_2(a)$ correction in Eq.~(\ref{eq:I2diag}).
 While the lattice errors in the two-boson system are nearly linear in $a$ over a fairly wide range, the higher powers of $a$ become 
stronger as we increase the number of bosons.
This goes hand-in-hand with the decreasing radius of the bound state, and  is a signature that we are probing higher-momentum 
scales as we increase the number of particles. 

\begin{figure}[!t]
\center
\includegraphics[width=1.0\textwidth]{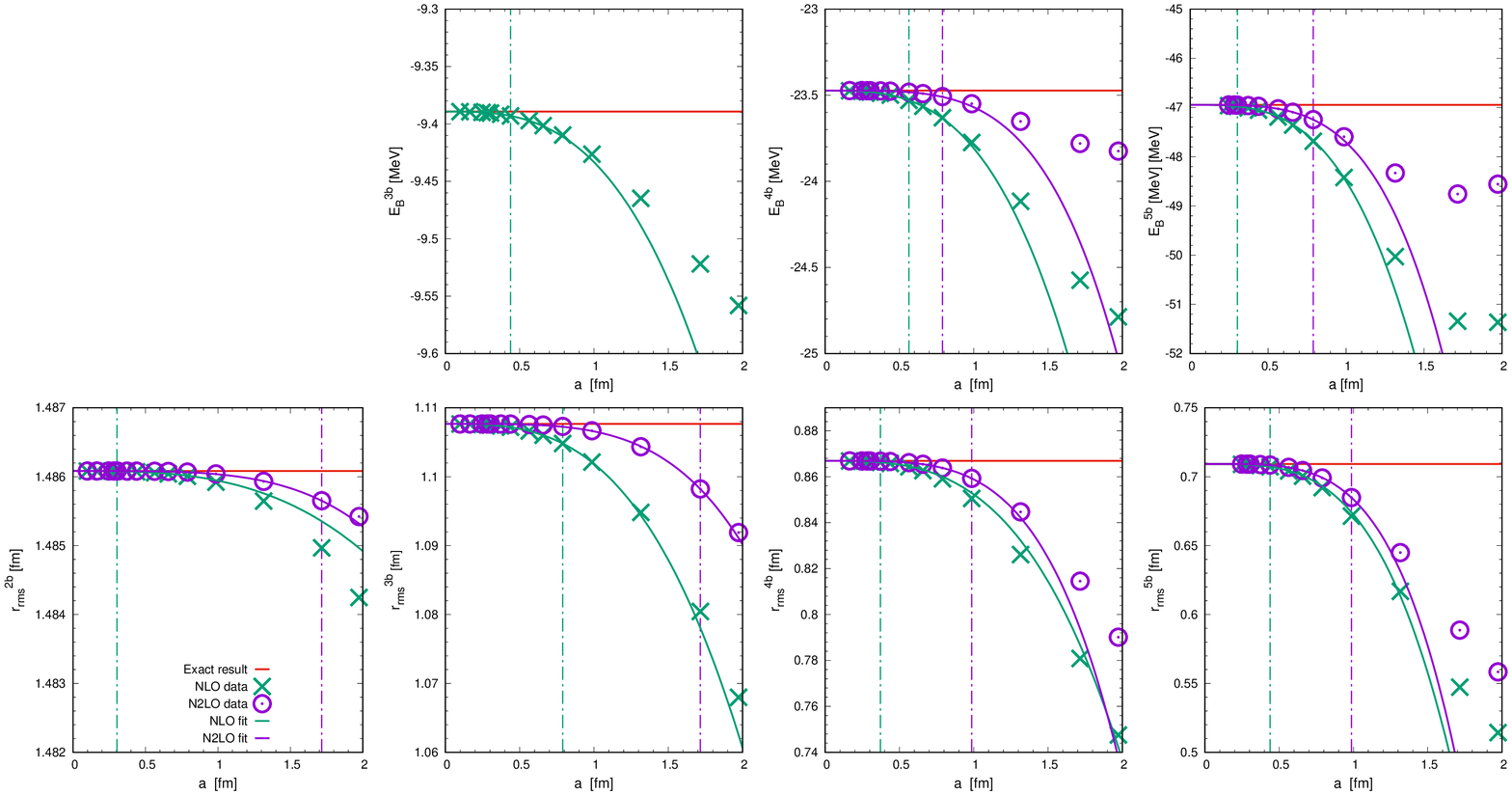}
\caption{Bound-state energies and root-mean-square radius versus lattice spacing for up to five bosons at NLO and N2LO. 
The vertical lines give the upper limits of the fit range.}
\label{fig:energyrmsN2LO}
\end{figure}

The NLO and N2LO bound state energies
and root-mean-square radii for up to five boson are shown in Fig.~\ref{fig:energyrmsN2LO} as well as the best fits using the asymptotic 
form in Eq.~(\ref{eq:extrap}).
 The crosses indicate the NLO results while the open circles are the N2LO results. The vertical lines gives the
upper limits of the fit range.  As shown in rows one through eight of Table~\ref{tab:obssum}, the NLO exponent $b_O$ is about 3 while 
the  N2LO exponent $b_O$ is about 4. This is consistent with our $\mathcal{O}(a^3)$ estimate
of errors at NLO and $\mathcal{O}(a^4)$ estimate of
errors at N2LO.
 We again see that the higher powers of $a$ become stronger as
we increase the number of bosons, and this is likely responsible for some systematic errors in our fitted values for $b_O$ in the five-boson system.

\begin{figure}[!t]
\center
\includegraphics[width=0.25\textwidth]{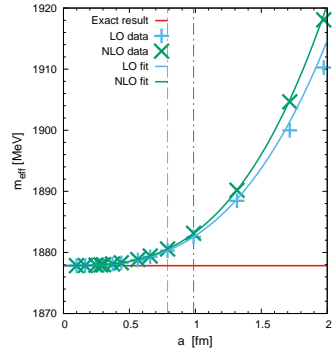}
\caption{Effective mass in the two-boson sector at LO and NLO. The vertical lines gives the upper limits of the fit range.}
\label{fig:meff2N}
\end{figure}

The lattice results for the effective two-boson mass at LO and NLO are shown
Fig.~\ref{fig:meff2N} as well as the best fits using Eq.~(\ref{eq:extrap}).  We omit the N2LO results since they are directly
fit to the continuum-limit results.
As shown in row nine of Table~\ref{tab:obssum}, the values for $b_O$ for each case is about 3.  This is consistent with our perturbative
analysis for the dependence on the total momentum.  We found that the
function $B_2(a)$ in Eq.~(\ref{eq:I2diag}) scales as $\mathcal{O}(a^3)$. 

\begin{figure}[!t]
\center
\includegraphics[width=0.5\textwidth]{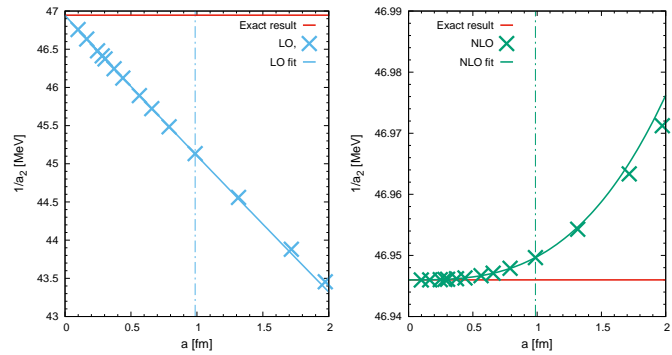}
\caption{The two-boson  inverse scattering length $1/a_2$ versus lattice spacing at LO and NLO.  The dotted
vertical lines give the upper limits of the fit range.}
\label{fig:1+1scattering}
\end{figure}

\begin{figure}[!t]
\center
\includegraphics[width=0.25\textwidth]{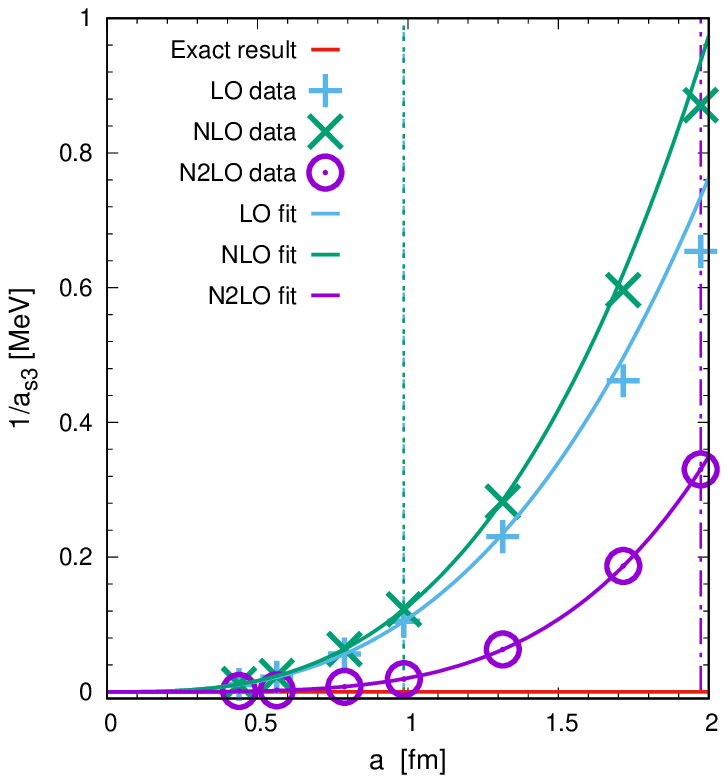}
\caption{The dimer-boson inverse  scattering length $1/a_3$ versus lattice spacing
at LO, NLO, and N2LO.  The vertical lines give the upper limits of the fit range.}
\label{fig:2+1scattering}
\end{figure}

In Fig.~\ref{fig:1+1scattering} we show the two-boson inverse scattering length $1/a_2$ at LO and NLO. The difference between the lattice scattering length at N2LO and the continuum value is so small that residual errors such as
finite-volume effects get in the way of a systematic analysis of lattice spacing dependence. The dotted
vertical lines give the upper limits of the fit range.  As shown in row ten of Table~\ref{tab:obssum}, the values 
for $b_O$ are about 1 at LO and about 3 at NLO.  These are consistent with our $\mathcal{O}(a^1)$ estimate
of errors at LO and $\mathcal{O}(a^3)$ estimate
of errors at NLO.  In Fig.~\ref{fig:2+1scattering} we show the dimer-boson inverse scattering length
at LO, NLO, N2LO. The dotted
vertical lines give the upper limits of the fit range.  As shown in row eleven
of Table~\ref{tab:obssum}, the values for $b_O$ is about 3 at LO,
3 at NLO, and 4 at N2LO.  These are consistent with our $\mathcal{O}(a^3)$ estimate of errors at NLO and $\mathcal{O}(a^4)$ estimate
of errors at N2LO.  The smaller than expected errors at LO can be understood from the fixed-point behavior of the dimer-boson inverse 
scattering length $1/a_3$.  As can be seen in Eq.~(\ref{eq:ptan3}), $1/a_3$ remains zero independent of the value of $1/a_2$.  As a result 
the $\mathcal{O}(a^1)$ error in $1/a_3$ vanishes, and the LO\ errors start only at   $\mathcal{O}(a^3)$. 

\section{Summary and outlook}\label{sec:summary}
In this paper we have applied the Symanzik lattice improvement program to lattice calculations
of zero-range bosons in one spatial dimensions.  We have considered up to five-boson bound states and computed bound-state energies, 
root-mean-square radii, effective masses, and inverse scattering lengths for two-boson and dimer-boson scattering.  For these 
calculations we have constructed the lattice action at LO, NLO, and N2LO and demonstrated that the size of the lattice errors are 
consistent with a perturbative analysis of cutoff effects in individual Feynman diagrams.  We have found that at LO the lattice errors 
are $\mathcal{O}(a^1)$, unless  suppressed due to kinematical reasons or special fixed-point behavior, as noted in a couple of examples above.  
Meanwhile, the errors at NLO are
typically $\mathcal{O}(a^3),$ and the errors at N2LO are $\mathcal{O}(a^4)$. The removal of any of the  terms in the improved action
would result in a larger lattice error in the continuum limit.  

The N2LO three-boson interaction proportional to $C_{\rm 3b}$ is particularly interesting.  We have used it to cancel lattice errors 
at $\mathcal{O}(a^3)$, and it does indeed remove residual errors in the structure and binding of bound states with three or more bosons.  
We note that for zero-range bosons in three spatial dimensions, the three-boson interaction is required at LO to properly renormalize the theory 
\cite{Efimov:1970zz,Bedaque:1998km}, see also \cite{Epelbaum:2016ffd}.
  While a four-boson contact interaction is not needed for renormalization in the three-dimensional theory \cite{Platter:2004he,Platter:2004ns}, 
one lesson learned from the analysis presented here is that the four-boson interaction could be useful as part of a comprehensive 
lattice improvement program to remove lattice errors in systems with four or more bosons.         

The lattice improvement program for more complicated systems in lattice effective field theory can proceed along similar lines, but may 
require more effort.  Some complications occur in effective field theories where non-perturbative iteration is used and singular interactions 
prevent the limit of infinite cutoff momentum.  For example, this is the case in chiral effective field theory for nucleons (see Ref.~\cite{Epelbaum:2008ga} 
for a review). In such cases one would need to verify the convergence of improved lattice formulations over a finite range of lattice 
spacings with each other rather than comparing with the limit of zero lattice spacing.  Similar issues strategies are used in non-relativistic QCD 
for heavy quarks on the lattice, where the lattice spacing should not be much smaller than the Compton wavelength of the heavy 
quark \cite{Lepage:1992tx,Davies:1997mg,Spitz:1999tu}. 

While the analysis we presented here is for a simple system of zero-range bosons in one dimension, it  serves as a useful guide for 
increasing the accuracy of more complicated calculations
in lattice effective field theory with improved lattice actions.
For zero-range bosons in one dimension, we found that our estimates of the lattice errors from naive dimensional analysis 
of individual diagrams were accurate.  However, this is directly related to the ultraviolet finiteness of our one-dimensional system.  
In general, we will need to regulate and renormalize divergent diagrams.  And, therefore, the renormalization group will be needed to compute 
anomalous dimensions that modify the naive dimensional analysis.    

\section*{Acknowledgments}
NK thanks S.~Elhatisari for useful discussions and carefully reading the manuscript. We acknowledge partial
financial support
from the CRC110: Deutsche Forschungsgemeinschaft (SGB/TR 110, ``Symmetries
and the Emergence of Structure in QCD''),
and the U.S. Department of Energy (DE-SC0018638).
Further support
was provided by the Chinese Academy of Sciences (CAS) President's International
Fellowship Initiative (PIFI) (grant no. 2018DM0034) and by VolkswagenStiftung (grant no. 93562).
The computational resources 
were provided by the J\"{u}lich Supercomputing Centre at Forschungszentrum
J\"{u}lich, RWTH Aachen, and the University of Bonn.

\begin{appendix}

\section{Loop integrals}\label{app:loops}
In the following we give more details about the terms appearing in Eq.~(\ref{eq:I2diag}) and Eq.~(\ref{eq:I3diag}). 
In our notation, $(E_i,p_i)$ are the incoming energies and momenta and $(E_i^\prime,p_i^\prime)$ are the outgoing energies
and momenta. The non-relativistic propagator for a single boson is defined as
\begin{equation}
i S(E,p)=\frac{i}{E-\frac{p^2}{2m}+i\epsilon},
\end{equation}
and the one-loop integral for the two-boson diagram in Fig.~\ref{I2} reads
\begin{equation}
\begin{split}
\mathcal{I}_2 & \left(E_1,p_1,E_2,p_2;E_1^\prime,p_1^\prime,E_2^\prime,p_2^\prime\right)
\sim\int_{\pi/a}^{\pi/a}\frac{dp}{2\pi} \int\frac{dp_0}{2\pi}S\left(E_{12}-p_0,p+p_{12}\right)S\left(p_0,p\right)\\
&\sim \frac{i m }{2} \sqrt{\frac{1}{-m E_{12}}}-\frac{1}{16\sqrt{m}}\left(\frac{1}{E_{12}}\right)^\frac{3}{2}p_{12}^2
-\frac{i m a}{\pi ^2}-\frac{i E_{12} m^2}{3 \pi ^4}a^3+\frac{m p_{12}^2}{6 \pi ^4}a^3+\mathcal{O}(a^5,p^4),
\end{split}
\label{eq:I2full}
\end{equation}
where we use energy and momentum conservation, $E_{12}=E_1+E_2=E_1^\prime+E_2^\prime$ and $p_{12}=p_1+p_2=p_1^\prime+p_2^\prime$, respectively. 
The $A_2(a)$,~$B_2(a)$ and $C_2(a)$ can be read off immediately from Eq.~(\ref{eq:I2full}).

The one-loop integral for the three-boson diagram in Fig.~\ref{I3} is given by 
\begin{equation}
\begin{split}
\mathcal{I}_3 & \left(E_1,p_1,E_2,p_2,E_3,p_3;E_1^\prime,p_1^\prime,E_2^\prime,p_2^\prime,E_3^\prime,p_3^\prime\right)\sim\int_{\pi/a}^{\pi/a}\frac{dp}{2\pi}\int\frac{dp_0}{2\pi}S\left(E_{12}-p_0,p+p_{12}\right)S\left(p_0,p\right)S\left(p_0,p\right)\\
&\sim\frac{-1}{4}\frac{\sqrt{m}}{E_{12}^{\frac{3}{2}}}-\frac{3}{32}p_{12}^2\frac{\sqrt{m}}{E_{12}^{\frac{5}{2}}}+\frac{i m^2}{3\pi^4}a^3+\mathcal{O}(a^5,p^4),
\end{split}\label{eq:I3}
\end{equation}
where we choose for simplicity that $E_3=E_3^\prime=0$, $p_3=p_3^\prime=0$ and set $E_{12}=E_1+E_2=E_1^\prime+E_2^\prime$ and $p_{12}=p_1+p_2=p_1^\prime+p_2^\prime$. 
Then $A_3(a)$ is given by Eq.~(\ref{eq:I3}) for this choice of momenta and energies.

\end{appendix}

\end{document}